\documentclass[doublecol]{epl2}

\title{Information filtering via hybridization of similarity preferential diffusion processes}
\usepackage{amssymb}
\usepackage{xcolor}
\author{An Zeng, Alexandre Vidmer, Matus Medo and Yi-Cheng Zhang\footnote{yi-cheng.zhang@unifr.ch}}
\shortauthor{An Zeng et. al.}

\institute{Department of Physics, University of Fribourg, Chemin du Mus\'{e}e 3, CH-1700 Fribourg, Switzerland\\
}
\pacs{89.75.-k}{Complex systems}
\pacs{89.65.-s}{Social and economic systems}
\pacs{89.20.Ff}{Computer science and technology}

\abstract{The recommender system is one of the most promising ways to address the information overload problem in online systems. Based on the personal historical record, the recommender system can find interesting and relevant objects for the user within a huge information space. Many physical processes such as the mass diffusion and heat conduction have been applied to design the recommendation algorithms. The hybridization of these two algorithms has been shown to provide both accurate and diverse recommendation results. In this paper, we proposed two similarity preferential diffusion processes. Extensive experimental analyses on two benchmark data sets demonstrate that both recommendation and accuracy and diversity are improved duet to the similarity preference in the diffusion. The hybridization of the similarity preferential diffusion processes is shown to significantly outperform the state-of-art recommendation algorithm. Finally, our analysis on network sparsity show that there is significant difference between dense and sparse system, indicating that all the former conclusions on recommendation in the literature should be reexamined in sparse system.}

\begin{document}

\maketitle

\section{Introduction}
The development of internet and World Wide Web bring many websites that allow
a large number of users to interact and share information. Examples include the well-known \emph{Twitter.com}, \emph{facebook.com} and so on. Day by day, these
online systems are rapidly growing, leading to a massive amount of available information for the users. The accessible information is for sure far more than every individual's ability to deal with, which is usually refereed as the information overload problem. Therefore, the information
filtering techniques nowadays becomes a very necessary and useful tool for online users. Recommendation is one the filtering techniques with the highest potential and widest application~\cite{Adomavicius05,Cacheda11}. Generally, it predicts users' taste and interested objects based on their historical records.

So far, various kinds of algorithms have been proposed, including collaborative filtering approaches~\cite{Herlocker04}, content-based analyses~\cite{Balabanovic1997}, tag-aware algorithms~\cite{Zhangzk2010}, trust-aware algorithms~\cite{Burke2007} and social impact based algorithms~\cite{ZengWEPJB2013}. For a review in this field, see ref.~\cite{PRlinyuan2012}. Recently, the information filtering has attracted more and more attention from physicists. Some classic diffusion processes in physics have been introduced to design recommendation algorithms. The mass diffusion algorithm enjoys a very high recommendation accuracy~\cite{Zhou07} while the heat conduction algorithm can generate very personalized recommendation results~\cite{Zhang07}. The hybrid approach of these two algorithms can achieve both high recommendation accuracy and diversity~\cite{Zhou10}.

Based on the hybrid method in~\cite{Zhou10}, many extensions have been made. For example, the setting of initial configuration~\cite{LiuC2012}, adding of the ground node~\cite{ZhouY2013}, personalized hybrid parameters~\cite{GuanY2013} are shown to further improve the recommendation performance. Moreover, the network manipulation has been shown to effectively solve the cold-start problem in recommendation~\cite{ZhangFG2012}. To enhance the efficiency of the recommendation process, the method to extract the information backbone (minimum structure) from online system is also designed~\cite{ZhangQM2013}. Very recently, the long-term influence of the recommendation methods on the user-item bipartite network evolution is studied~\cite{EPL9718005}. It is found that many personalized recommendation methods have reinforce effect on item degree distribution in long term.

Different from the above studies which focus on modifying the way to hybrid the diffusion processes or manipulating the underlying bipartite networks, we directly improve the basic diffusion process by introducing the similarity preference mechanism. Actually, the concept of degree preferential diffusion has been used in information filtering~\cite{PRE83066119,PRE84037101}. By enhancing the diffusion resource on small degree objects, these degree preferential diffusion algorithms can recommend many niche and novel objects. Besides novelty, there is in fact another aspect of recommendation diversity called personalization which measures how the recommendation lists are different from user to user~\cite{PRlinyuan2012}. We find that the similarity preference mechanism can significantly improve the recommendation personalization of the mass diffusion and heat conduction algorithms. Moreover, the recommendation accuracy and novelty are increased accordingly. The hybridization of the similarity preferential diffusion processes is shown to significantly outperform the well-known method in~\cite{Zhou07}. Finally, our analysis on network sparsity show that there is significant difference between dense and sparse system, indicating that all the former conclusions on recommendation in the literature should be reexamined in sparse system.

\section{Recommendation algorithms}

\begin{figure}
  \center
  \includegraphics[width=\columnwidth]{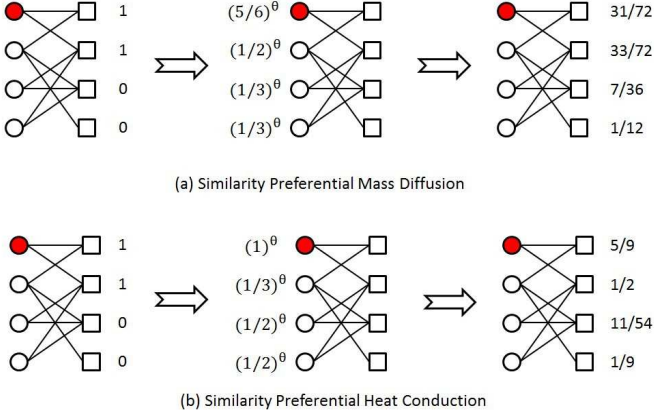}
\caption{(Color online) The illustration of the (a) preferential mass diffusion and (b) preferential heat conduction processes. Users are shown as circles; objects are squares. The target user is indicated by the shaded circle. $\theta$ is set as $2$ as an example here.}\label{fig1}
\end{figure}

The online commercial system can be modeled by a bipartite network, where users and objects are characterized by two distinct kinds of nodes. The bipartite network can be represented by
an adjacency matrix $A$, where the element $a_{i\alpha}$ equals $1$ if user $i$ has collected object $\alpha$, and $0$ otherwise. (throughout this paper objects are labeled by Greek letters, whereas users are identified by Latin letters).

We first describe the Similarity Preferential Mass Diffusion (SPMD) method.
For the target user $i$ to whom we recommend objects to, each of $i$'s collected object is assigned with one unit of
resource. The resource of each object is equally distributed to all the neighboring users who have collected this object. If user $j$ is one of these users, the resource he/she receives from object $\alpha$ will be $1/k_{\alpha}$ where $k_{\alpha}$ is degree of $\alpha$ (namely the number of users who collected $\alpha$). The final resource $j$ receives is the sum over all $i$'s collected objects:
\begin{equation}
f_{ij} = \sum_{\alpha=1}^{M}{\frac{a_{i\alpha}a_{j\alpha}}{k_{\alpha}}}.
\end{equation}

Actually, $f_{ij}$ can be used to measure the similarity between user $i$ and $j$.
From intuitive sense, the objects selected by more similar user $j$ should be more relevant to the target user $i$. We accordingly modify $f_{ij}$ after the second diffusion step as $f_{ij}^{\theta}$ where $\theta$ is a tunable parameter. When $\theta=1$, the method reduces to the classic mass diffusion process ~\cite{Zhou07}. When $\theta>1$, the user $j$ more similar to the target user $i$ will play a more important role in the following diffusion. In the last step of diffusion, we let each user distribute their resource $f_{ij}^{\theta}$ equally to the neighboring objects. The final resource object $\beta$ obtained is
\begin{equation}
f_{i\beta}=\sum_{j=1}^{N}\frac{a_{j\beta}f_{ij}^{\theta}}{k_j}
\end{equation}
where $k_j$ is the number of objects $j$ collected. The final resources of all objects will be sorted in descending order to generate the recommendation list for user $i$. The SPMD process is illustrated in Fig. 1(a). From the community structure point of view, this modification means that the objected selected by the users in the same communities are more likely to be recommended to the target user. A similar study of using the community structure information to improve the link prediction performance can be found in~\cite{Yanb2012}.

The Similarity Preferential Heat Conduction (SPHC) method works similar to the SPMD algorithm, but instead follows diffusion diffusion formulas as
\begin{equation}
f_{ij} = \sum_{\alpha=1}^{M}{\frac{a_{i\alpha}a_{j\alpha}}{k_{j}}}.
\end{equation}
$f_{ij}^{\theta}$ is used in the final step of diffusion as SPMD as
\begin{equation}
f_{i\beta}=\sum_{j=1}^{N}\frac{a_{j\beta}f_{ij}^{\theta}}{k_\beta}.
\end{equation}
When $\theta=1$, the SPHC method degenerates to the classic Heat conduction process \cite{Zhang07}. The SPHC process is illustrated in Fig. 1(b).

Finally, we consider the nonlinear hybridization of the SPMD and SPHC algorithms. The formula for the first step of diffusion reads as
\begin{equation}
f_{ij} = \sum_{\alpha=1}^{M}{\frac{a_{i\alpha}a_{j\alpha}}{k_{\alpha}^{\lambda}k_{j}^{1-\lambda}}}.
\end{equation}
The formula for the second step of diffusion is
\begin{equation}
f_{i\beta}=\sum_{j=1}^{N}\frac{a_{j\beta}f_{ij}^{\theta}}{k_j^{\lambda}k_{\beta}^{1-\lambda}}.
\end{equation}
The parameter $\lambda$ adjusts the relative weight between the two algorithms. When $\lambda$ increases from $0$ to $1$, the hybrid algorithm changes gradually from SPHC to SPMD. When $\theta=1$, such hybrid approach is exactly the same as the method proposed in ref.~\cite{Zhou10}.

\section{Data}
To test the performance of the algorithm, we make use of two benchmark
data sets. The MovieLens data is freely available in~\cite{movielens}. It
consists of 1682 movies and 943 users who can rate the movies from 1 to 5 (i.e., the worst to the best). The original data contains $10^5$ ratings. We consider only the ratings higher than 2 as a link here. After such filtering, the data contains 82520 user-item pairs. The Netflix data~\cite{netflix} is a random sampling of the
whole records of user activities in \emph{Netflix.com}. It has 3000 users, 3000 movies. Similar to the
MovieLens data, only the links with ratings larger than 2 are
considered. Finally, the data has 197248 links. Each data is
randomly divided into two parts: the training set contains
90\% of the data ($E^T$) and the remaining 10\% of data constitutes the probe set ($E^P$). The recommendation algorithm will run on $E^T$ while $E^P$ will be used to estimate the recommendation performance.

\section{Metrics}
In order to measure the accuracy of the recommendation algorithm, we adopt
\emph{ranking score} ($RS$) index. Specifically, $RS$ measures whether the ranking of the items in the recommendation list matches the users' real taste. For each recommendation algorithm, it will provide each user with a ranking list of all his uncollected items. For a target user $i$, we calculate the position for each of his link in the probe set. If one of his uncollected item $\alpha$ is ranked at the $3$th place and the total number of his uncollected items is $100$, $RS_{i\alpha}=3/100=0.03$. In an accurate recommendation, the items in the probe set should be ranked higher, corresponding to a smaller $RS$. As such, the mean value of the $RS$ over all the links in the probe set can be used to evaluate the recommendation accuracy as
\begin{equation}
RS=\frac{1}{|E^{P}|}\sum_{i\alpha\in E^{P}}RS_{i\alpha}.
\end{equation}
According to the definition, a well-performed recommendation algorithm should have small $RS$.

In real online systems, users are normally provided with only the top part of the recommendation list. We will use another more practical recommendation accuracy measurement called \emph{precision}, which is only based on each user's top-$L$ items in the recommendation list. For a target user $i$, his precision of recommendation is calculated as
\begin{equation}
P_{i}(L)=\frac{d_{i}(L)}{L},
\end{equation}
where $d_{i}(L)$ represents the number of user $i$'s probe set links contained in
the top-$L$ recommendation list. The precision $P(L)$ for the whole system can be obtained by simply averaging the precisions
over all users with at least one link in the probe set.

Actually, the information need of a user usually goes beyond several best sellers in the online systems. Since predicting users' personalized preference is much more difficult, the diversity in the recommendation list has been recognized as another crucial criteria in judging the recommendation results besides accuracy. In this letter, we make use of two kinds of diversity measurement: \emph{personalization} and \emph{novelty}.

\begin{figure}
  \center
  \includegraphics[width=\columnwidth]{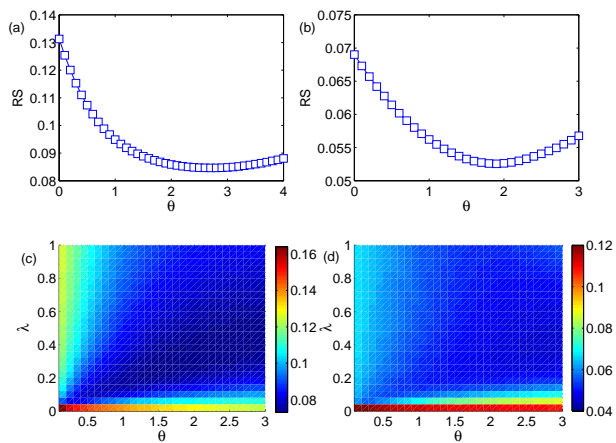}
\caption{(Color online) Dependence of the ranking score of the SPMD method on parameter $\theta$ for (a) movielens and (b) netflix. (c) and (d) are the ranking score of the hybrid method of SPMD and SPHC in parameter space ($\theta$, $\lambda$) for movielens and netflix, respectively. }\label{fig2}
\end{figure}

\begin{figure*}
  \center
  \includegraphics[width=18cm]{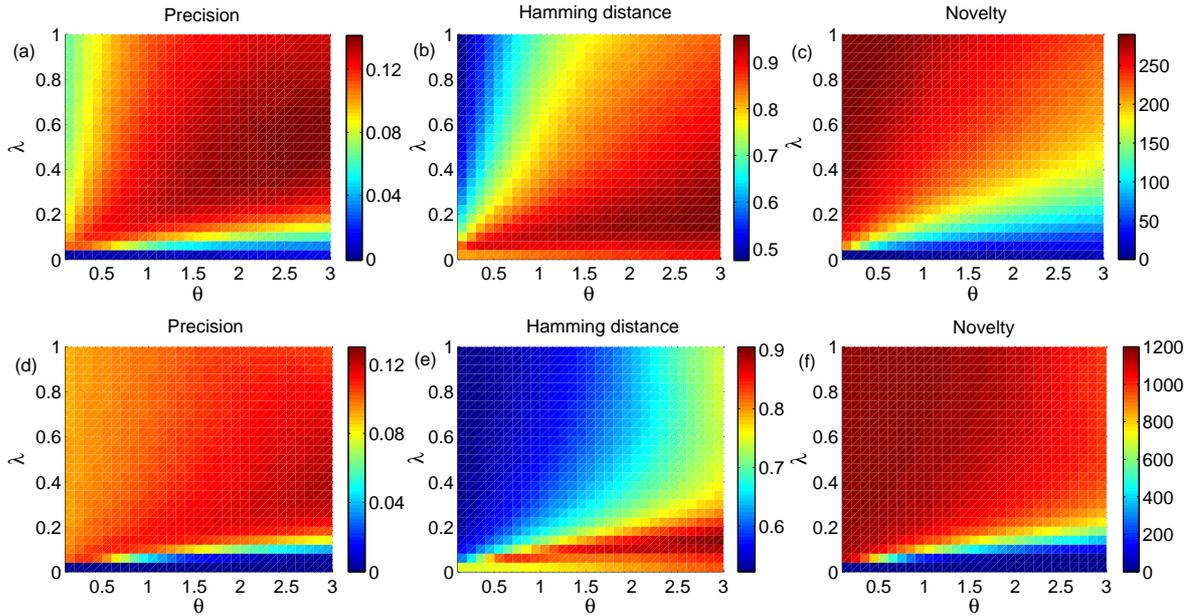}
\caption{(Color online) The precision $P$ in ($\theta$, $\lambda$) plane for (a) movielens and (d) netflix. The hamming distance $H$ in ($\theta$, $\lambda$) plane for (b) movielens and (e) netflix. The novelty $N$ in ($\theta$, $\lambda$) plane for (c) movielens and (f) netflix.}\label{fig3}
\end{figure*}

The personalization mainly consider how users' recommendation lists are different from one to another. Normally, it is measured by the Hamming distance. Denoting $C_{ij}(L)$ as the number of overlapped items in the top-$L$ place of the recommendation list of user $i$ and $j$, their hamming distance can be calculated as
\begin{equation}
H_{ij}(L)=1-\frac{C_{ij}(L)}{L}.
\end{equation}
$H_{ij}(L)$ is between $0$ and $1$, which are corresponding to the cases where these two users have the same or entirely different recommendation lists, respectively. To estimate the personalization of a recommendation algorithm, we calculate the mean hamming distance $H(L)$ by averaging $H_{ij}(L)$ over all pairs of users. A higher $H(L)$ indicates a more personalized recommendation.

The novelty measures the average degree of the items in the recommendation list. For those popular items, users can easily get them from many other channels. However, it's generally difficult for the users to find the unpopular item they are interested in. Therefore, a good recommender system should include a reasonable number of small degree items in the recommendation list. The \emph{novelty} index can be defined as
\begin{equation}
N_{i}(L)=\frac{1}{L}\sum_{\alpha\in O^{i}}k_{\alpha},
\end{equation}
where $O^{i}$ denotes the recommendation list for user $i$. A highly novel and unexpected recommendation of items results in a small mean $N(L)$.

\section{Results}

We start from investigating the performance of the SPMD method. The results are presented in Fig. 2(a) and (b). An minimum ranking score $RS$ can be easily observed in both networks and the optimal $\theta$s are larger than $1$ ($\theta^*=2.6$ in movielens and $\theta^*=1.9$ in netflix). The $RS$s are improved by $12.42\%$ in movielens and $6.41\%$ in netflix, respectively.

The performance of the hybrid method of the SPMD and SPHC method is studied as well. The dependence of $RS$ on parameter $\lambda$ and $\theta$ is shown as heatmap in Fig. 2(c) and (d). Actually, the hybrid method of MD and HC algorithm is shown to achieve a better $RS$ than pure MD and pure HC~\cite{Zhou10}. We observe here that a minimum $RS$ still exists when $\theta>1$ and the optimal $RS$ under all possible value of $\lambda$ and $\theta$ happens in the region where $\theta>1$. Specifically, the optimal parameters are $\theta^*=2$ and $\lambda^*=0.32$ in movielens. In netflix, the optimal parameters are $\theta^*=3$ and $\lambda^*=0.4$. Compared to the original hybrid method of MD and HC, the $RS$s are improved by $5.04\%$ in movielens and $12.02\%$ in netflix. These results indicate that introducing the similarity preference in the diffusion processes can indeed enhance the recommendation accuracy.

We then report the precision $P(L)$, hamming distance $H(L)$ and novelty $N(L)$ of the hybrid method in Fig. 3. All these metrics depend on the recommendation list length $L$. In this paper, we set $L=20$ according to the literature~\cite{Zhou10}. The results of $P(L)$ in Fig. 3(a) and (d) confirm our finding that the preferential diffusion can improve the recommendation accuracy of the hybrid method. By using the optimal parameters determined from the $RS$, the $P(L)$ is improved by $8.99\%$ in movielens and $10.49\%$ in netflix, compared to the original hybrid method.

\begin{table*}[htbp]
\begin{center}

\caption{The results of all the metrics for different recommendation algorithms.}
\label{Table1}
\begin{tabular}{cccc cccc cccc cccc cccc cccc cccc}
\\
\hline
\hline
Network & Method & Ranking score &Precision &Hamming distance &Novelty\\
\hline
            & MD & 0.0958  & 0.1146 & 0.7030 & 278.1\\
            & SPMD  & 0.0839 & 0.1286 & 0.8367 & 237.6\\
Movielens   & HC & 0.1351 & 0.0063 & 0.8620 & 6.53\\
            & SPHC  & 0.1216 & 0.0134 & 0.9039 & 13.85\\
            & Hybrid (MD+HC)   & 0.0754 & 0.1291 & 0.9025 & 178.8\\
            & Hybrid (SPMD+SPHC)    & 0.0716 & 0.1407 & 0.9171 & 174.5\\

 \hline
           & MD  & 0.0562  & 0.0990 & 0.5508 & 1169\\
           & SPMD  & 0.0526  & 0.1055 & 0.6316 & 1116\\
Netflix    & HC  & 0.1125  & 0.0003 & 0.7630 & 1.35\\
           & SPHC  & 0.1084  & 0.0005 & 0.8298 & 2.16\\
           & Hybrid (MD+HC)  & 0.0516 & 0.1115 & 0.6669& 1058\\
           & Hybrid (SPMD+SPHC)  & 0.0454 & 0.1232& 0.7592 & 955\\

\hline
\hline
\end{tabular}
\vspace*{0.0cm}
\end{center}
\end{table*}

In addition to accuracy, the recommendation diversity is of great significance.
For personalization, we can estimate how the recommendation results are different from user to user. A larger hamming distance indicates a more personalized recommendation. Besides personalization, the novelty is also an important aspect. With a small novelty, the average degree of the recommended items are low, so that more fresh items will appear in the recommendation list. From Fig. 3 (b) and (e), it can be seen that not only the hybrid parameter $\lambda$ can control the value of hamming distance, the influence of $\theta$ is significant as well. Specifically, the hamming distance increase with $\theta$. This is because the most similar users are different from one user to another and the similarity preferential diffusion amplify the weight of the most similar users, leading to a more personalized recommendation at the end. We can see in Fig. 3(c) and (f) that the novelty is enhanced by the parameter $\theta$ too. By using the optimal parameters determined from the $RS$, $H$ and $N$ are improved by $1.62\%$ and $2.40\%$ in movielens, respectively. In netflix,  $H$ and $N$ are improved by $13.84\%$ and $9.74\%$, respectively.

The detailed results of all the metrics above are listed in Table I. It is already well-known that heat conduction algorithm can generate a very diverse recommendation and is considered to be one of the most diverse recommendation algorithms so far. Interestingly, the SPHC method can further improve the recommendation diversity of the original HC method. As shown in Table I, its hamming distance is significantly increased.

We further investigate the effects of data sparsity on the algorithmic performance. For the whole data set, we select a fraction $1-p$ ($p$ ranging from 0.1 to
0.9 with step 0.1) links as the training set; the fraction
$p$ of the links form the training set. Clearly, lower $p$ indicates sparser data (i.e., less information). Here, we mainly focus on the recommendation accuracy (measured by ranking score).
We report the minimum $RS^*$ of the hybrid method of SPMD and SPHC, and the corresponding optimal parameters under different setting of $p$ in Fig. 4. Obviously, the hybrid method of SPMD and SPHC enjoys a lower $RS^*$ than the original hybrid method in both data sets.

Moreover, there are some interesting phenomenon in the optimal parameters. In the inset of Fig. 4, we can see that the hybrid parameter $\lambda^*$ keeps increasing with the data sparsity $p$, indicating the SPMD method should take a more important role in the hybrid method. This is natural because SPMD inclines to recommend popular objects and it is generally safer to do so when there is very few historical information of each user. Unlike $\lambda^*$, $\theta^*$ keeps decreasing with $p$, indicating personalization is less and less important when data is sparse. When $p>0.8$, $\theta^*$ is even smaller than $1$, which means the diffusion should consider more the less similar users. In other words, there is critical difference between dense systems and sparse systems when applying recommendation algorithms. Since most of real systems are sparse, these results suggest that we should recheck in sparse data about all the conclusions on recommendation based on the dense data.

\begin{figure}
  \center
  \includegraphics[width=\columnwidth]{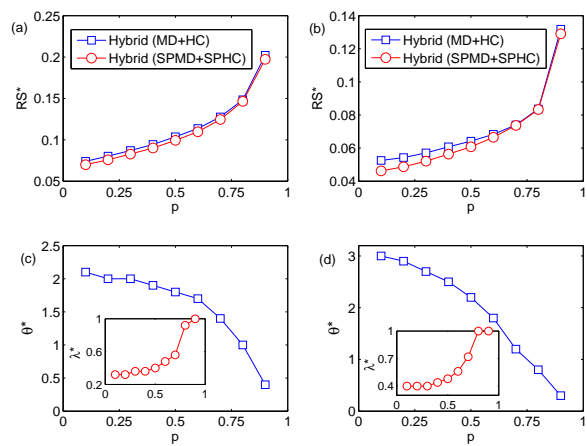}
\caption{(Color online) The minimum $RS^*$ of two different hybrid methods under different $p$ in (a) movielens and (b) netflix. The optimal parameters of the hybrid method of SPMD and SPHC with respect to $RS*$ under different $p$ in (c) movielens and (d) netflix.}\label{fig4}
\vspace{-0.1cm}
\end{figure}

\begin{table*}[htbp]
\begin{center}

\caption{The results of all the metrics for different recommendation algorithms under the triple data division.}
\label{Table1}
\begin{tabular}{cccc cccc cccc cccc cccc cccc cccc}
\\
\hline
\hline
Network & Method & Ranking score &Precision &Hamming distance &Novelty\\
\hline

Movielens   & Hybrid (MD+HC)   & 0.0809  & 0.1133  & 0.8771  & 178.5 \\
            & Hybrid (SPMD+SPHC)    & 0.0781  & 0.1198  & 0.9041 & 167.7 \\

 \hline

Netflix    & Hybrid (MD+HC)  &0.0536  &0.1025  &0.6107 &996.1 \\
           & Hybrid (SPMD+SPHC)  &0.0482  &0.1125 &0.7452  &862.1 \\

\hline
\hline
\end{tabular}
\vspace*{0.0cm}
\end{center}
\end{table*}

How to choose the parameters in the recommendation algorithms is an important issue in practice. If the optimal parameters vary significantly over time in real systems, the recommendation algorithm might not be meaningful from practical point of view. To test our algorithm in this aspect, we consider the triple division of the data. Specifically, the data is
randomly divided into three parts: the training set contains 80\% of the data, another 10\% forms the testing set and the remaining 10\% of data constitutes the probe set. Both the training set and testing set are treated as known data (``historical data") and the testing set is used to estimate the optimal parameters for the recommendation algorithm. We run the recommendation algorithm on the training set and choose the parameters when the recommendation accuracy ($RS$) in the testing set is optimized. The parameters will be considered as the optimal parameters to apply to the ``future" (the probe set). We compare the hybrid method of SPMD and SPHC (with two parameters: $\lambda$ and $\theta$) to the hybrid method of MD and HC (with one parameter $\lambda$). The results can be seen in table II. Obviously, even though our method has one more parameter, the recommendation performance in both accuracy and diversity is better than the hybrid method of MD and HC.

\section{Conclusion}
In this paper, we proposed the similarity preferential mass diffusion (SPMD) and similarity preferential heat conduction (SPHC) processes. In the preferential diffusion, the effect of similar users will be amplified, so that the user can receive more personalized recommendations. Interestingly, the recommendation accuracy can be improved as well due the similarity preference in the diffusion. Moreover, we hybrid the SPMD and SPHC algorithms and we find that it can remarkably outperform the original hybrid method~\cite{Zhou10} in both recommendation accuracy and diversity. We finally investigate the effect of network sparsity on our algorithms. Even though our hybrid method can constantly outperforms the original hybrid method, the parameter changes significantly under different sparsity setting. Since the properties of sparse data are essential different from dense data, some recommendation algorithms specific for sparse systems should be designed. We finally test our method in triple data division and our method can still outperform the original hybrid. These results show that the optimal parameters are very stable in each data sets and support the effectiveness of our method when applied to real systems.

This work can lead to many applications. For example, the link prediction in directed networks depends on the so-called ``Bi-fan" structure~\cite{ZhangQM20132}. The similarity preference can be introduced as a weighting strategy in this structure and improve the prediction precision. More generally, such similarity preferential diffusion can be applied to any multi-step of diffusion process on networks. There are actually many different global and local diffusion based methods to estimate the similarity between nodes~\cite{PRlinyuan2012,ZengA2011}. We believe that the similarity preference mechanism can improve them.

\acknowledgments
This work was partially supported by the Future and Emerging Technologies program
of the European Commission FP7-COSI-ICT (project QLectives, grant no.
231200) and by the Swiss National Science Foundation (grant no. 200020-143272).

\end{document}